\documentclass[11pt]{article}
\usepackage[T1]{fontenc}
\usepackage[latin9]{inputenc}
\usepackage[a4paper]{geometry}
\usepackage[active]{srcltx}
\usepackage{amsmath}
\usepackage{amssymb}
\usepackage{esint}

\makeatletter
\pdfoutput=1 

\usepackage{jheppub}

\usepackage{etoolbox}
    
    \patchcmd{\maketitle}{\@fpheader}{}{}{}

\usepackage{amsfonts}

\usepackage{bbm}

\title{\boldmath On conserved charges and thermodynamics of the AdS$_{4}$ dyonic black hole}

\author[a,b]{Marcela C\'ardenas,}
\author[a]{Oscar Fuentealba}
\author[a]{and Javier Matulich}

\affiliation[a]{Centro de Estudios Cient\'{i}ficos (CECs), Av. Arturo Prat 514, Valdivia,
Chile.}
\affiliation[b]{Departamento de F\'isica, Universidad de Concepci\'on, Casilla 160-C, Concepci\'on, Chile.}

\emailAdd{cardenas@cecs.cl}
\emailAdd{fuentealba@cecs.cl}
\emailAdd{matulich@cecs.cl}

\abstract{Four-dimensional gravity in the presence of a dilatonic scalar field and an Abelian gauge field is considered. This theory corresponds to the bosonic sector of a Kaluza-Klein dimensional reduction of eleven-dimensional supergravity which induces a determined self-interacting potential for the scalar field. We compute the conserved charges and carry out the thermodynamics of an anti-de Sitter (AdS) dyonic black hole solution recently proposed. The charges coming from symmetries of the action are computed by using the Regge-Teitelboim Hamiltonian approach. These correspond to the mass, which acquires contributions from the scalar field, and the electric charge. Integrability conditions are introduced because the scalar field leads to non-integrable terms in the variation of the mass. These conditions are generically solved by introducing boundary conditions that arbitrarily relates the leading and subleading terms of the scalar field fall-off. The Hamiltonian Euclidean action, computed in the grand canonical ensemble, is obtained by demanding the action to attain an extremum. Its value is given by a radial boundary term plus an additional polar angle boundary term due to the presence of a magnetic monopole. Remarkably, the magnetic charge can be identified from the variation of the additional polar angle boundary term, confirming that the first law of black hole thermodynamics is a consequence of having a well-defined and finite Hamiltonian action principle, even if the charge does not come from a symmetry of the action. The temperature and electrostatic potential are determined demanding regularity on the black hole solution, whereas the value of the magnetic potential is already identified in the variation of the additional polar angle boundary term. Consequently, the first law of black hole thermodynamics is identically satisfied by construction.}

\makeatother

\begin{document}
\maketitle \flushbottom

\newpage{}

\section{Introduction}

It has been recently conjectured about the need of additional terms
in the first law of black hole thermodynamics, interpreted as scalar
charges for a new class of dyonic black hole introduced in \cite{Lu:2013ura},
\cite{Lu:2014maa}. This black hole is asymptotically AdS provided
the system is endowed with a self-interacting potential. The theory
is described by the action of the bosonic sector coming from a consistent
truncation of a $S^{7}$ reduction of eleven-dimensional supergravity
\cite{Cvetic:1999xp}. In the thermodynamic analysis of \cite{Lu:2013ura},
the authors claimed that the first law is not satisfied unless one
adds a term which is interpreted as a scalar charge \cite{Lu:2013ura},
\cite{Lu:2014maa}. However, this argument conflicts with the fact
that there is no gauge symmetry related to the presence of a single
scalar field, i.e., there is no a Noether charge associated to this
Lagrangian (nor a topological charge as is the case of the magnetic
charge). Moreover, it has been clearly identified that the scalar
field generically contributes to the mass with non-integrable terms.
This has been proved with general asymptotic conditions through Hamiltonian
\cite{Henneaux:2006hk} and other methods \cite{Barnich:2002pi},
\cite{Compere:2009dp}, and even with an explicit black hole example
in presence of gauge fields in three dimensions \cite{Cardenas:2014kaa}.
In this context, in \cite{Chow:2013gba} is presented a new class
of dyonic AdS black hole solutions of four-dimensional $\mathcal{N}=8$,
SO$(8)$ gauge supergravity where the AdS$_{4}$ dyonic dilatonic
black hole of \cite{Lu:2013ura} is included. It is found in fact
that the missing term, supposedly related to a scalar charge, is a
contribution to the variation of the mass. It is also pointed out
in \cite{Chow:2013gba} that the non-integrability of this term leads
to an ill-defined mass with independent electric and magnetic charges.
However, as it will be shown in this manuscript, it is possible to
impose general integrability conditions. One can settle some physical
criterium that is manifested through suitable boundary conditions
which determine the arbitrary functions coming from the integrability
conditions. One possible criterium is to demand the preservation of
the AdS symmetry of the scalar field fall-off \cite{Hertog:2004dr},
\cite{Henneaux:2006hk}, obtaining that the scalar field contribution
is cancelled by a term coming from the gravitational contribution
to the mass.

Other motivation for studying the black hole solution presented in
\cite{Lu:2013ura} has to do with the computation of its Gibbs free
energy. For doing so, it requires to have a well defined action and
also presenting other thermodynamic features such that to exhibit
all the charges of the solution with their respective chemical potentials.
The Gibbs free energy presented in \cite{Lu:2013ura} fails on that,
since they cannot recover the magnetic contribution to the Euclidean
action, not even including the supposed scalar charge. The value of
the action in \cite{Lu:2013ura} was obtained through the holographic
renormalization method described in \cite{deHaro:2000vlm} by adding
counterterms which only considers radial surface terms to get a finite
action principle. As we will show bellow, it is necessary an additional
polar angle boundary term for obtaining the magnetic contribution
to the Euclidean action. For proving the latter, we formulate a well-defined
and finite Hamiltonian action principle for the system and we prove
that this additional boundary term comes from a total derivative in
the polar angle which appears due to the black hole solution is endowed
with a magnetic monopole.

The aim of this paper is to compute the conserved charges of the AdS$_{4}$
dyonic black hole and also to formulate a well-defined and finite
Hamiltonian action principle for the described system which allows
to obtain the value of the Hamiltonian Euclidean action. By imposing
the grand canonical ensemble, the Gibbs free energy is chosen as our
thermodynamic potential which, unlike the free energy computed in
\cite{Lu:2013ura}, exhibits all the conserved charges of the solution,
i.e., the mass (with its respective scalar field contribution), the
electric charge and the magnetic charge. The plan of this manuscript
is the following, in the Section \ref{Section2} we present the Lagrangian
and the AdS$_{4}$ dyonic dilatonic black hole solution of \cite{Lu:2013ura}.
Section \ref{Section3} is focused on the Hamiltonian analysis
and the corresponding conserved charges. The mass and the electric
charge are computed by using the Regge-Teitelboim Hamiltonian approach.
In the variation of the mass there are two contributions, the gravitational
and the scalar field part (already identified in \cite{Chow:2013gba}).
Integrability conditions are needed to be imposed since the presence
of the scalar field leads to a non-integrable term. Suitable boundary
conditions are chosen in order to preserve the AdS symmetry of the
scalar field fall-off. This implies a precise relation on the coefficients
of the leading and subleading terms of the scalar field, as was noted
for this kind of systems in \cite{Henneaux:2006hk}. In Section \ref{Section4}
we perform the thermodynamic analysis of the solution. In there,
the Hamiltonian Euclidean action is introduced, where for simplicity
the calculations are done in a suitable Euclidean minisuperspace.
For obtaining the Gibbs free energy we compute the value of the Euclidean
Hamiltonian action endowed with a suitable radial boundary term and
an additional polar angle boundary term. These terms are needed to
be added in order to have a well-defined and finite Hamiltonian action
principle. From the variation of the boundary term at infinity is
possible to identify the variation of the Hamiltonian conserved charges
of this system, which are the mass and electric charge. On the other
hand, the variation of the magnetic charge comes from the additional
polar boundary term. This boundary term has to be considered due to
the presence of a magnetic monopole. The chemical potentials associated
to Noether charges are Lagrange multipliers of the system at infinity.
They are obtained through regularity conditions at the horizon, unlike
the magnetic potential which remarkably already appears determined
in the variation of the boundary term besides the magnetic charge.
It is worth to note that the first law of black hole thermodynamics
is satisfied independently of the integrability conditions on the
mass, since the relation only involves the variation of the conserved
charges. Once the Gibbs free energy is obtained, the value of the
mass, electric charge, magnetic charge and entropy are verified
by using the known thermodynamic relations. Finally, Section \ref{Section5}
is devoted to some concluding remarks.

\section{AdS$_{4}$ dyonic black hole solution \label{Section2}}

We consider four-dimensional gravity with negative cosmological constant
in presence of an Abelian gauge field and a dilatonic scalar field
with a self-interacting potential. The action reads
\begin{equation}
I[g_{\mu\nu},A_{\mu},\phi]=\int d^{4}x\sqrt{-g}\left(\frac{R}{2\kappa}-\frac{1}{2}g^{\mu\nu}\partial_{\mu}\phi\partial_{\nu}\phi-\frac{1}{4}e^{-\sqrt{3}\phi}F^{\mu\nu}F_{\mu\nu}-V\left(\phi\right)\right).\label{eq:Action1}
\end{equation}
Hereafter the gravitational constant is chosen as $\kappa=1/2$.%
\footnote{The vacuum permeability constant located in front of the Maxwell-like
action in \eqref{eq:Action1} turns out to be normalized to one after
the dimensional reduction.%
} The self-interacting potential of the scalar field is given by
\begin{equation}
V\left(\phi\right)=-6g^{2}\cosh\left(\frac{\phi}{\sqrt{3}}\right),\label{eq:Self-Int-Pot}
\end{equation}
where the coupling constant $g$ determines the AdS radius as $\ell^{2}=g^{-2}$.
The theory given by \eqref{eq:Action1} corresponds to the bosonic
sector of two possible dimensional reductions, which depend on the
coupling constant $g$. In the case of vanishing $g$, the action
is obtained after a $S^{1}$ reduction of five-dimensional pure gravity.
On the other hand if $g\neq0$ the action can be obtained after a
$S^{7}$ reduction of eleven-dimensional supergravity as was found
in \cite{Cvetic:1999xp}.

The gravitational field equations for the action \eqref{eq:Action1}
are
\begin{equation}
G_{\mu\nu}=T_{\mu\nu}^{\phi}+T_{\mu\nu}^{A},
\end{equation}
where the contributions to the energy-momentum tensor of the dilatonic
scalar field and the gauge field are given by
\begin{eqnarray}
T_{\mu\nu}^{\phi} & = & \frac{1}{2}\partial_{\mu}\phi\partial_{\nu}\phi-\frac{1}{4}g_{\mu\nu}\partial^{\lambda}\phi\partial_{\lambda}\phi+\frac{1}{2}g_{\mu\nu}V(\phi),\\
T_{\mu\nu}^{A} & = & \frac{1}{2}e^{-\sqrt{3}\phi}\left(F_{\mu}^{\,\,\,\lambda}F_{\nu\lambda}-\frac{1}{4}g_{\mu\nu}F^{\lambda\rho}F_{\lambda\rho}\right),
\end{eqnarray}
respectively. The equation for the scalar field is 

\begin{equation}
\Box\phi+\frac{\sqrt{3}}{4}e^{-\sqrt{3}\phi}F^{\mu\nu}F_{\mu\nu}-\frac{dV}{d\phi}=0,
\end{equation}
and the equation for the gauge field reads

\begin{equation}
\nabla_{\mu}\left(e^{-\sqrt{3}\phi}F^{\mu\nu}\right)=0.
\end{equation}

This system admits an AdS dyonic black hole which is static and spherically
symmetric \cite{Lu:2013ura}. The line element of this configuration
can be written as 
\begin{equation}
ds^{2}=-\left(H_{1}H_{2}\right)^{-1/2}fdt^{2}+\frac{dr^{2}}{\left(H_{1}H_{2}\right)^{-1/2}f}+\left(H_{1}H_{2}\right)^{1/2}r^{2}\left(d\theta^{2}+\sin^{2}\left(\theta\right)d\varphi^{2}\right),\label{eq:BHLor}
\end{equation}
where the functions determining the solution are given by
\begin{equation}
f\left(r\right)=f_{0}\left(r\right)+g^{2}r^{2}H_{1}\left(r\right)H_{2}\left(r\right),\quad f_{0}\left(r\right)=1-\frac{2\mu}{r},\label{eq:function-f}
\end{equation}
\begin{equation}
H_{1}\left(r\right)=\gamma_{1}^{-1}\left(1-2\beta_{1}f_{0}\left(r\right)+\beta_{1}\beta_{2}f_{0}\left(r\right)^{2}\right),\quad H_{2}\left(r\right)=\gamma_{2}^{-1}\left(1-2\beta_{2}f_{0}\left(r\right)+\beta_{1}\beta_{2}f_{0}\left(r\right)^{2}\right),
\end{equation}
with $\gamma_{1}=1-2\beta_{1}+\beta_{1}\beta_{2}$, and $\gamma_{2}=1-2\beta_{2}+\beta_{1}\beta_{2}$.
The dilatonic scalar field is given by 
\begin{equation}
\phi\left(r\right)=\frac{\sqrt{3}}{2}\log\left(\frac{H_{2}\left(r\right)}{H_{1}\left(r\right)}\right),\label{eq:Scalar-Field}
\end{equation}
whereas the one-form gauge field has the following form
\begin{equation}
A=A_{t}(r)dt+A_{\varphi}(\theta)d\varphi.\label{eq:Alor}
\end{equation}
The time component of \eqref{eq:Alor} is 
\begin{equation}
A_{t}\left(r\right)=\frac{\sqrt{2}\left(1-H_{1}\left(r\right)-\beta_{1}\left(f_{0}-H_{1}\left(r\right)\right)\right)}{\sqrt{\beta_{1}\gamma_{2}}H_{1}\left(r\right)},
\end{equation}
while the definition of the angular component of the gauge potential
changes depending on the hemisphere in order to avoid the Dirac string
\cite{Dirac:1948um}. Hence
\begin{equation}
A_{\varphi}\left(\theta\right)=\begin{cases}
p(1+\cos\left(\theta\right)) & ,\quad0\leq\theta<\frac{\pi}{2}-\delta,\\
p\left(-1+\cos\left(\theta\right)\right) & ,\quad\frac{\pi}{2}+\delta<\theta\leq\pi,
\end{cases}\label{eq:Aphi}
\end{equation}
where $p=2\sqrt{2}\mu\gamma_{2}^{-1}\sqrt{\beta_{2}\gamma_{1}}$ and
$\delta\rightarrow0$ (Wu-Yang monopole \cite{Wu:1976qk}, \cite{Wu:1976ge}).
In this solution the coordinates range as $0<r<\infty$, $-\infty<t<\infty$,
$0\leq\theta<\pi$ and $0\leq\varphi<2\pi$. All the integration constants
($\mu$, $\beta_{1}$, $\beta_{2}$ , $\gamma_{1}$, $\gamma_{2}$)
are restricted to be positive.

In the case of $\beta_{1}=\beta_{2}$, the dilatonic scalar field
is decoupled and the solution turns out to be an AdS dyonic Reissner-Nordström
black hole where the electric and magnetic charges have the same value.
If $\beta_{1}=0$ the solution is purely magnetic and in the case
of $\beta_{2}=0$ the configuration becomes purely electric. If $\mu=0$
the solution turns out to be the anti-de Sitter spacetime.

\section{Hamiltonian and surface integrals\label{Section3}}

The Hamiltonian generator for the Lagrangian\eqref{eq:Action1} reads
as
\begin{equation}
H\left[\xi,\xi^{A}\right]=\int d^{3}x\left(\xi^{\perp}\mathcal{H}_{\perp}+\xi^{i}\mathcal{H}_{i}-\xi^{A}\mathcal{G}\right)+Q\left[\xi,\xi^{A}\right],\label{eq:Ham-Gen}
\end{equation}
where the boundary term $Q\left[\xi,\xi^{A}\right]$, which corresponds
to the conserved charges in the Regge-Teitelboim approach, ensures
that the Hamiltonian generator has well-defined functional derivatives
\cite{Regge:1974zd}. The bulk term appearing in \eqref{eq:Ham-Gen}
is a linear combination of the constraints $\mathcal{H}_{\perp}$,
$\mathcal{H}_{i}$ and $\mathcal{G}$, where the first two are the
energy and momentum densities, and the last one corresponds to the
Gauss constraint associated to the Abelian gauge field. The asymptotic
surface deformations of the spacetime are given by the vector $\xi=\left(\xi^{\perp},\xi^{i}\right)$,
and $\xi^{A}$ in turn is the gauge parameter of the Abelian symmetry.
The constraints are explicitly given by
\begin{eqnarray}
\mathcal{H}_{\perp} & = & \frac{1}{\sqrt{\gamma}}\left(\pi^{ij}\pi_{ij}-\frac{1}{2}\left(\pi_{\,\, i}^{i}\right)^{2}\right)-\sqrt{\gamma}R\nonumber \\
 &  & +\frac{\pi_{\phi}^{2}}{2\sqrt{\gamma}}+\sqrt{\gamma}\left(\frac{1}{2}\partial^{i}\phi\partial_{i}\phi+V(\phi)\right)+e^{\sqrt{3}\phi}\frac{\pi^{i}\pi_{i}}{2\sqrt{\gamma}}+\frac{1}{4}\sqrt{\gamma}e^{-\sqrt{3}\phi}F^{ij}F_{ij},\\
\mathcal{H}_{i} & = & 2\nabla_{j}\pi_{\,\, i}^{j}+\pi_{\phi}\partial_{i}\phi+\pi^{j}F_{ij},\\
\mathcal{G} & = & \partial_{i}\pi^{i}.
\end{eqnarray}
The dynamical variables of the system are the spatial components of
the fields $\left\{ \gamma_{ij},\, A_{i},\,\phi\right\} $, where
$\gamma_{ij}$ is the spatial metric of the ADM decomposition. Here
$R$ stands for the scalar curvature of the 3-dimensional spatial
metric $\gamma_{ij}$ and the self-interacting potential of the scalar
field $V\left(\phi\right)$ is defined in eq. \eqref{eq:Self-Int-Pot}.
The momentum conjugated to the 3-dimensional metric $\gamma_{ij}$
is 
\begin{equation}
\pi^{ij}=-\sqrt{\gamma}\left(K^{ij}-\gamma^{ij}K\right),
\end{equation}
where the extrinsic curvature is given by
\begin{equation}
K_{ij}=\frac{1}{2N^{\perp}}\left(\nabla_{i}N_{j}+\nabla_{j}N_{i}-\dot{\gamma}_{ij}\right).
\end{equation}
The momentum for the dilatonic field $\phi$ reads
\begin{equation}
\pi_{\phi}=\frac{\sqrt{\gamma}}{N^{\perp}}\left(\dot{\phi}-N^{i}\partial_{i}\phi\right),
\end{equation}
and for the gauge field $A_{i}$,
\begin{equation}
\pi^{i}=-\frac{\sqrt{\gamma}e^{-\sqrt{3}\phi}}{N^{\perp}}\left(-\gamma^{ij}F_{0j}+N^{j}\gamma^{ik}F_{jk}\right).
\end{equation}

The variation of the surface term exhibits different contributions
according to the field content of the theory, such that
\begin{equation}
\delta Q\left[\xi,\xi^{A}\right]=\delta Q^{G}\left[\xi,\xi^{A}\right]+\delta Q^{\phi}\left[\xi,\xi^{A}\right]+\delta Q^{A}\left[\xi,\xi^{A}\right],
\end{equation}
where $\delta Q$ was obtained after demanding that $\delta H=0$
on the constraint surface (vanishing constraints). The explicit expressions
for the surface integrals are given by

\begin{eqnarray}
\delta Q^{G} & = & \intop dS_{l}G^{ijkl}\left(\xi^{\perp}\nabla_{k}\delta\gamma_{ij}-\partial_{k}\xi^{\perp}\delta\gamma_{ij}\right)+\intop dS_{l}\left[2\xi_{k}\delta\pi^{kl}+\left(2\xi^{k}\pi^{jl}-\xi^{l}\pi^{kj}\right)\delta\gamma_{jk}\right],\nonumber \\
\\
\delta Q^{\phi} & = & -\intop dS_{i}\left(\xi^{\perp}\sqrt{\gamma}\partial^{i}\phi\delta\phi+\xi^{i}\pi_{\phi}\delta\phi\right),\\
\delta Q^{A} & = & -\intop dS_{i}\left[\xi^{\perp}\sqrt{\gamma}e^{-\sqrt{3}\phi}F^{ij}\delta A_{j}+\left(\xi^{i}\pi^{j}-\pi^{j}\xi^{i}\right)\delta A_{j}-\xi^{A}\delta\pi^{i}\right],
\end{eqnarray}
with 
\begin{equation}
G^{ijkl}=\frac{1}{2}\sqrt{\gamma}\left(\gamma^{ik}\gamma^{jl}+\gamma^{il}\gamma^{jk}-2\gamma^{ij}\gamma^{kl}\right).
\end{equation}

\subsection{Conserved charges of the AdS$_{4}$ dyonic black hole}

In order to obtain the above surface integrals, let us consider a
static and spherically symmetric minisuperspace in which the AdS$_{4}$
dyonic black hole \eqref{eq:BHLor} is included. For simplicity in
the upcoming analysis we perform the following change of variable
in the radial coordinate 
\begin{equation}
\rho^{2}=\sqrt{H_{1}\left(r\right)H_{2}\left(r\right)}r^{2}.
\end{equation}
The line element then reads
\begin{equation}
ds^{2}=-N^{\perp}\left(\rho\right)^{2}dt^{2}+\frac{d\rho^{2}}{F\left(\rho\right)}+\rho^{2}\left(d\theta^{2}+\sin^{2}\left(\theta\right)d\varphi^{2}\right).
\end{equation}
The gauge field ansatz is given by
\begin{equation}
A=A_{t}\left(\rho\right)dt+A_{\varphi}\left(\theta\right)d\varphi,
\end{equation}
and the scalar field also depends on the radial coordinate $\phi=\phi(\rho)$.
Taking in consideration this minisuperspace the only nonvanishing
momentum is the radial component of the electromagnetic one, where
$\pi^{\rho}=p^{\rho}\left(\rho,\theta\right)$. Therefore, the value
of the Hamiltonian charges, computed on the sphere $S^{2}$ of infinite
radius, is given by
\begin{eqnarray}
\delta Q & = & \left[-\xi^{t}\left(\frac{8\pi\rho N^{\perp}\delta F}{\sqrt{F}}+4\pi\sqrt{F}N^{\perp}\rho^{2}\partial_{\rho}\phi\delta\phi+\pi\left[\left(\int\frac{N^{\perp}e^{-\sqrt{3}\phi}}{\sqrt{F}\rho^{2}}\, d\rho\right)\csc(\theta)\delta A_{\varphi}\partial_{\theta}A_{\varphi}\right]_{\theta=0}^{\theta=\pi}\right)\right.\nonumber \\
 &  & +\left.2\pi\xi^{A}\intop_{0}^{\pi}\delta p^{\rho}d\theta\right]_{\rho\rightarrow\infty}.\label{eq:delta carga}
\end{eqnarray}
Here, we have applied the definition of the deformation vectors $\xi^{\perp}$
and $\xi^{i}$ in terms of the Killing vectors $\xi^{t}$ and $\bar{\xi}^{i}$,
which reads
\begin{eqnarray}
\xi^{\perp} & = & N^{\perp}\xi^{t},\\
\xi^{i} & = & \bar{\xi}^{i}+N^{i}\xi^{t}.
\end{eqnarray}
In order to compute and perform a proper analysis of the charges,
we must give suitable asymptotic conditions that represent the behavior
of the fields at infinity. These conditions are specified up to the
orders that contribute to the charges, such that
\begin{eqnarray}
F\left(\rho\right) & = & g^{2}\rho^{2}+1+F_{0}+\frac{F_{1}}{\rho}+\mathcal{O}\left(\frac{1}{\rho^{2}}\right),\\
N^{\perp}\left(\rho\right) & = & g\rho+\mathcal{O}\left(\frac{1}{\rho}\right),\\
\phi\left(\rho\right) & = & \frac{\phi_{1}}{\rho}+\frac{\phi_{2}}{\rho^{2}}+\mathcal{O}\left(\frac{1}{\rho^{3}}\right),\\
p^{\rho}\left(\rho,\theta\right) & = & p_{0}\sin\left(\theta\right)+\mathcal{O}\left(\frac{1}{\rho^{1}}\right),\\
\xi^{A} & = & \xi_{0}^{A}+\mathcal{O}\left(\frac{1}{\rho}\right).
\end{eqnarray}
The coefficients in the expansions given above are parameters that
depend on the integration constants of the corresponding solution.
The variation of the charge obtained after replacing the proposed
asymptotic behavior in \eqref{eq:delta carga} is given by%
\footnote{It has to be noted that a divergent term appears in the variation
of the charge but it vanishes once the solution is replaced. This
is because the divergent part of the gravitational contribution is
cancelled by the divergent part of the scalar field contribution by
virtue of the relation $\delta F_{0}=\frac{g^{2}}{2}\phi_{1}\delta\phi_{1}$.%
}

\begin{equation}
\delta Q=\xi^{t}\left[-8\pi\delta F_{1}+4\pi g^{2}\left(2\phi_{2}\delta\phi_{1}+\phi_{1}\delta\phi_{2}\right)\right]+4\pi\xi_{0}^{A}\delta p_{0}.\label{eq:deltacarga2}
\end{equation}
The mass is the conserved charge associated to the time spacetime
translations, which in this approach is obtained from $\delta M=\delta Q\left[\xi^{t}\right]$,
while the electric charge is the charge associated to the Abelian
gauge transformations, where $\delta Q_{e}=\delta Q\left[\xi^{A}\right]$.
Then, the variation of the mass and the electric charge reads
\begin{align}
\delta M & =-8\pi\delta F_{1}+4\pi g^{2}\left(2\phi_{2}\delta\phi_{1}+\phi_{1}\delta\phi_{2}\right),\label{eq:dM0}\\
\delta Q_{e} & =4\pi\delta p_{0},
\end{align}
respectively. The electric charge can be directly integrated for the
AdS$_{4}$ dyonic black hole, which in terms of the integration constants
of the solution is written as 
\begin{equation}
Q_{e}=\frac{16\pi\sqrt{2}\mu\sqrt{\beta_{1}\gamma_{2}}}{\gamma_{1}}.\label{eq:Q}
\end{equation}
In contrast, the mass is generically non-integrable and its variation
is explicitly given by
\begin{equation}
\delta M=\delta\left(\frac{16\pi\left(1+\beta_{1}\right)\left(1-\beta_{2}\right)\left(1-\beta_{1}\beta_{2}\right)\mu}{\gamma_{1}\gamma_{2}}+\frac{64\pi g^{2}\mu^{3}\left(1-\beta_{1}\beta_{2}\right)\left(\beta_{1}-\beta_{2}\right)^{2}\gamma}{\gamma_{1}^{3}\gamma_{2}^{3}}\right)+\Phi,\label{eq:deltaM}
\end{equation}
with
\begin{equation}
\gamma=\beta_{1}+\beta_{2}-8\beta_{1}\beta_{2}+6\beta_{1}^{2}\beta_{2}+6\beta_{1}\beta_{2}^{2}-8\beta_{1}^{2}\beta_{2}^{2}+\beta_{1}^{3}\beta_{2}^{2}+\beta_{1}^{2}\beta_{2}^{2}
\end{equation}
Note that the variation of the mass coincides with the one computed
in \cite{Chow:2013gba}, which has the non-integrable term $\Phi$
that comes from the scalar field part of the energy density. This
term is given by
\begin{equation}
\Phi=4\pi g^{2}\left(2\phi_{2}\delta\phi_{1}+\phi_{1}\delta\phi_{2}\right),
\end{equation}
where the leading and subleading terms of the scalar field fall-off
are respectively
\begin{eqnarray}
\phi_{1} & = & \frac{2\sqrt{3}\left(\ensuremath{\beta_{2}}\left(1+\beta_{1}{}^{2}\right)-\ensuremath{\beta_{1}}\left(1+\beta_{2}{}^{2}\right)\right)\mu}{\gamma_{1}\gamma_{2}},\\
\phi_{2} & = & \frac{2\sqrt{3}\left(-\text{\ensuremath{\beta}}_{2}^{2}\left(1-\ensuremath{\beta_{1}^{4}}\right)-2\text{\ensuremath{\beta}}_{1}\text{\ensuremath{\beta}}_{2}^{2}(-4+3\text{\ensuremath{\beta}}_{2})-2\beta_{1}^{3}\text{\ensuremath{\beta}}_{2}(-3+4\text{\ensuremath{\beta}}_{2})-\beta_{1}{}^{2}\left(-1+8\text{\ensuremath{\beta}}_{2}-8\text{\ensuremath{\beta}}_{2}^{3}+\text{\ensuremath{\beta}}_{2}^{4}\right)\right)\mu^{2}}{\gamma_{1}^{2}\gamma_{2}^{2}}.\nonumber \\
\end{eqnarray}

The presence of a non-integrable term $\Phi$ in the variation of
the mass \eqref{eq:dM0} demands integrability relations among the
fall-off coefficients of the scalar field. The integrability condition
$\delta^{2}M=0$ implies that $\delta\Phi=0$, this implies in turn
that $\phi_{2}=\phi_{2}\left(\phi_{1}\right)$. Hence, the mass generically
takes the form
\begin{equation}
M=-8\pi F_{1}+4\pi g^{2}\int\left(2\phi_{2}+\phi_{1}\frac{d\phi_{2}}{d\phi_{1}}\right)d\phi_{1}.\label{eq:Masss}
\end{equation}
In this point it is necessary to impose a boundary condition that
fixes a precise relation between the leading and subleading terms
of the scalar field behavior at infinity. One possible criterium is
to demand preservation of the AdS symmetry of the scalar field asymptotic
fall-off, which can be carried out since the AdS$_{4}$ dyonic dilatonic
black hole of \cite{Lu:2013ura} is within the asymptotic conditions
for AdS spacetimes analyzed in \cite{Hertog:2004dr}, \cite{Henneaux:2006hk}.
In there, they construct a set of boundary conditions for having a
well-defined and finite Hamiltonian generators for all the elements
of the anti-de Sitter algebra in the case of gravity minimally coupled
to scalar fields. Then, we are allowed to impose certain relations
on the leading and subleading terms of the scalar field fall-off provided
the scalar field does not break the AdS symmetry at infinity. These
boundary conditions are $\left(\phi_{1}=0,\quad\phi_{2}\neq0\right)$,
$\left(\phi_{1}\neq0,\quad\phi_{2}=0\right)$ and $\phi_{2}=c\phi_{1}^{2}$,
where $c$ is a constant without variation. In terms of the integration
constants the latter relation becomes
\begin{align}
-2(\beta_{1}-\beta_{2})\mu^{2}\left[-\left(\sqrt{3}+6c\right)\beta_{2}-\left(\sqrt{3}-6c\right)\beta_{1}{}^{3}\beta_{2}{}^{2}\right.\nonumber \\
-\text{\ensuremath{\beta}}_{1}\left(\sqrt{3}-6c-8\sqrt{3}\beta_{2}+6\left(\sqrt{3}-2c\right)\beta_{2}{}^{2}\right)\nonumber \\
\left.-\beta_{1}{}^{2}\beta_{2}\left(6\left(\sqrt{3}+2c\right)-8\sqrt{3}\beta_{2}+\left(\sqrt{3}+6c\right)\text{\ensuremath{\beta_{2}}}{}^{2}\right)\right] & =0.\label{eq:cond2}
\end{align}
From eq. \eqref{eq:cond2} we observe three cases, two of them nontrivial.
When $\mu=0$ \eqref{eq:cond2} holds, the mass, the electric charge
and magnetic charge vanish giving rise to the vacuum solution which
turns out to be the AdS$_{4}$ spacetime. The other two cases imply
that $\beta_{1}=\beta_{1}\left(\beta_{2}\right)$ in such a way that
they make the terms proportionals to $g^{2}$ in \eqref{eq:deltaM}
to vanish. Hence, the mass becomes the AMD mass \cite{Ashtekar:1984zz},
\cite{Ashtekar:1999jx} obtained in \cite{Lu:2013ura},
\begin{equation}
M=\frac{16\pi\left(1-\beta_{1}\right)\left(1-\beta_{2}\right)\left(1-\beta_{1}\beta_{2}\right)\mu}{\gamma_{1}\gamma_{2}}.
\end{equation}
This fact is in agreement with \cite{Anabalon:2014fla}, where it
is pointed out that in hairy spacetimes the mass does not receive
scalar field contributions provided the scalar field respects the
AdS invariance at infinity, making some holographic prescriptions
suitable for computing the mass.

\section{Thermodynamics of the AdS$_{4}$ dyonic black hole\label{Section4}}

The thermodynamic analysis of the AdS$_{4}$ dyonic dilatonic black
hole is performed in this section. We define the Euclidean Hamiltonian
action of the theory including a surface term and a additional polar
boundary term to have a finite action principle. The presence of the
latter additional term is due to the solution counts with a magnetic
monopole. For simplicity, we take a minisuperspace in which the AdS$_{4}$
dyonic black hole is included. The variation of the Euclidean Hamiltonian
action is computed in the grand canonical ensemble, where the chemical
potentials are fixed. Remarkably, the magnetic charge emerges from
the additional polar boundary term accompanied by its respective chemical
potential. The value of the temperature and electric potential, on
the other hand, are settled imposing regularity conditions. When the
variation of the additional surface and polar boundary terms is determined,
as was mentioned above, integrability conditions are needed to be
impose to determine the value of the Euclidean Hamiltonian action
leading to the Gibbs free energy.

\subsection{Hamiltonian action and Euclidean minisuperspace}

Let us consider spacetimes with manifold of topology $\mathbb{R}^{2}\times S^{2}$.
The plane $\mathbb{R}^{2}$ is centered at the event horizon $r_{+}$
and is parametrized by the periodic Euclidean time $\tau$ and the
radial coordinate $r$. These plane coordinates range as
\begin{gather}
0\leq\tau<\beta,\\
r_{+}\leq r<\infty,
\end{gather}
with $\beta$ the inverse of the Hawking temperature and the 2-sphere
$S^{2}$ stands for the topology of the base manifold. The Hamiltonian
Euclidean action for the system is given by
\begin{equation}
I^{E}=\intop_{0}^{\beta}d\tau\intop_{\Sigma}d^{3}x\left[\dot{\gamma}_{ij}\pi^{ij}+\dot{A}_{i}\pi^{i}+\dot{\phi}\pi_{\phi}-\left(N^{\perp}\mathcal{H}_{\perp}+N^{i}\mathcal{H}_{i}-A_{\tau}\mathcal{G}\right)\right]+B,\label{eq:HamI}
\end{equation}
where $\Sigma=\mathbb{R}\times S^{2}$ is the spatial section of the
manifold. Note that the additional term $B$ in \eqref{eq:HamI} needs
to be added to the action in order to have a well-defined variational
principle, and it is crucial for determining the value of the action
for stationary configurations.

The Euclidean continuation of the AdS$_{4}$ dyonic black hole \eqref{eq:BHLor}
is considered. The line element reads
\begin{equation}
ds^{2}=N^{\perp}\left(r\right)^{2}d\tau^{2}+\frac{dr^{2}}{F\left(r\right)}+H(r)\left(d\theta^{2}+\sin^{2}\left(\theta\right)d\varphi^{2}\right).\label{eq:Line-Element-MS}
\end{equation}
where the gauge field ansatz and the scalar field are given by
\begin{align}
A & =A_{\tau}\left(r\right)d\tau+A_{\varphi}\left(\theta\right)d\varphi,\\
\phi & =\phi(r).
\end{align}
The radial component of the electromagnetic field momentum is $\pi^{r}=p^{r}\left(r,\theta\right)$
(all the other momenta of the fields vanish). Hence, from \eqref{eq:HamI}
is possible to obtain the following reduced action
\begin{equation}
I^{E}=-2\pi\beta\intop_{r_{+}}^{\infty}dr\intop_{0}^{\pi}d\theta\left(N^{\perp}\left(r\right)\mathcal{H}_{\perp}-A_{\tau}\left(r\right)\mathcal{G}\right)+B,\label{eq:Red-Ac}
\end{equation}
where the reduced constraints take the form 
\begin{eqnarray}
\mathcal{H}_{\perp} & = & -\frac{e^{-\sqrt{3}\phi}\sin(\theta)}{2\sqrt{F}H}\left[-\csc^{2}(\theta)\left(\partial_{\theta}A_{\varphi}\right)^{2}-2e^{\sqrt{3}\phi}H\left(\partial_{r}F\partial_{r}H+2F\partial_{r}^{2}H-2\right)\right.\\
 &  & \left.+e^{\sqrt{3}\phi}H{}^{2}\left(12g^{2}\cosh\left(\frac{\phi}{\sqrt{3}}\right)-F\left(\partial_{r}\phi\right)^{2}\right)+e^{\sqrt{3}\phi}F\left(\partial_{r}H\right)^{2}+\csc^{2}(\theta)e^{2\sqrt{3}\phi}\left(p^{r}\right)^{2}\right],\nonumber \\
\mathcal{G} & = & \partial_{r}p^{r}.
\end{eqnarray}
The variation of the reduced action \eqref{eq:Red-Ac} with respect
to the Lagrange multipliers $N^{\perp}$ and $A_{\tau}$ indicates
that the constraints have to vanish 
\begin{equation}
\mathcal{H}_{\perp}=0,\qquad\mathcal{G}=0.\label{eq:Constraints}
\end{equation}
These equations define the constraint surface. On the other hand,
the variation of \eqref{eq:Red-Ac} with respect to the independent
functions of the dynamical fields in the minisuperspace leads to the
field equations. The field equations related to $F\left(r\right)$
and $H\left(r\right)$ are respectively given by 
\begin{align}
\frac{e^{-\sqrt{3}\phi}\sin(\theta)}{4F{}^{3/2}H}\left(N^{\perp}\left(-\csc^{2}(\theta)\left(\partial_{\theta}A_{\varphi}\right)^{2}-F\left(\partial_{r}H\right)^{2}e^{\sqrt{3}\phi}+\csc^{2}(\theta)e^{2\sqrt{3}\phi}\left(p^{r}\right)^{2}\right)\right.\nonumber \\
\left.+H{}^{2}N^{\perp}e^{\sqrt{3}\phi}\left(F\left(\partial_{r}\phi\right)^{2}+12g^{2}\cosh\left(\frac{\phi}{\sqrt{3}}\right)\right)+4He^{\sqrt{3}\phi}\left(N^{\perp}-F\partial_{r}H\partial_{r}N^{\perp}\right)\right) & =0,\label{eq:eq1}\\
\nonumber \\
\frac{e^{-\sqrt{3}\phi}\sin(\theta)}{2\sqrt{F}H{}^{2}}\left(N^{\perp}\left(-\csc^{2}(\theta)\left(\partial_{\theta}A_{\varphi}\right)^{2}-F\left(\partial_{r}H\right)^{2}e^{\sqrt{3}\phi}+\csc^{2}(\theta)e^{2\sqrt{3}\phi}\left(p^{r}\right)^{2}\right)\right.\nonumber \\
-H{}^{2}e^{\sqrt{3}\phi}\left(N^{\perp}\left(12g^{2}\cosh\left(\frac{\phi}{\sqrt{3}}\right)-F\left(\partial_{r}\phi\right)^{2}\right)-2\left(\partial_{r}F\partial_{r}N^{\perp}+2F\partial_{r}^{2}N^{\perp}\right)\right)\nonumber \\
\left.+He^{\sqrt{3}\phi}\left(N^{\perp}\left(\partial_{r}F\partial_{r}H+2F\partial_{r}^{2}H\right)+2F\partial_{r}H\partial_{r}N^{\perp}\right)\right) & =0.
\end{align}
The field equations associated to $A_{\varphi}\left(r,\theta\right)$
and $p^{r}\left(r,\theta\right)$ are
\begin{equation}
\frac{N^{\perp}e^{-\sqrt{3}\phi}\csc(\theta)}{\sqrt{F(r,s)}H(r,s)}\left(\partial_{\theta}A_{\varphi}\cot(\theta)-\partial_{\theta}^{2}A_{\varphi}\right)=0,\qquad\partial_{r}A_{\tau}+\frac{\csc(\theta)N^{\perp}e^{\sqrt{3}\phi}p^{r}}{\sqrt{F}H}=0,
\end{equation}
and finally the scalar field equation reads

\begin{align}
-\frac{e^{-\sqrt{3}\phi}\sin(\theta)}{2\sqrt{F}H}\left(N^{\perp}\left(\sqrt{3}\csc^{2}(\theta)\left(\partial_{\theta}A_{\varphi}\right)^{2}+2FHe^{\sqrt{3}\phi}\partial_{r}H\partial_{r}\phi\right.\right.\nonumber \\
+H{}^{2}e^{\sqrt{3}\phi}\left(\partial_{r}F\partial_{r}\phi+2F\partial_{r}^{2}\phi+4\sqrt{3}g^{2}\sinh\left(\frac{\phi}{\sqrt{3}}\right)\right)\nonumber \\
\left.\left.+\sqrt{3}\csc^{2}(\theta)e^{2\sqrt{3}\phi}\left(p^{r}\right)^{2}\right)+2FH{}^{2}e^{\sqrt{3}\phi}\partial_{r}\phi\partial_{r}N^{\perp}\right) & =0.\label{eq:eq5}
\end{align}
Then, the variation of the reduced action \eqref{eq:Red-Ac} on the
constraint surface and evaluated on-shell (eqs. from \eqref{eq:eq1}
to \eqref{eq:eq5} have to be satisfied) turns out to be

\begin{eqnarray}
\delta I^{E}\Big|_{on-shell} & = & -2\pi\beta\int_{0}^{\pi}d\theta\left[N^{\perp}\sin\left(\theta\right)\left(\frac{\partial_{r}H\delta F+\partial_{r}F\delta H}{\sqrt{F}}-\frac{\sqrt{F}\partial_{r}H\delta H}{H}\right.\right.\nonumber \\
 &  & \left.\left.+\frac{\sqrt{F}H\partial_{r}\phi\delta\phi}{2}+2\sqrt{F}\partial_{r}\delta H\right)-\partial_{r}\left(2N^{\perp}\sin\left(\theta\right)\sqrt{F}\right)\delta H-A_{\tau}\delta p^{r}\right]_{r_{+}}^{\infty}\nonumber \\
 &  & -2\pi\beta\int_{r+}^{\infty}dr\left[\frac{N^{\perp}e^{-\sqrt{3}\phi}}{H\sqrt{F}\sin\theta}\partial_{\theta}A_{\varphi}\delta A_{\varphi}\right]_{0}^{\pi}+\delta B.\label{eq:variationaction}
\end{eqnarray}
If we demand the action to attain an extremum, i.e., $\delta I^{E}\Big|_{on-shell}=0$,
the variation of additional term $\delta B$ necessarily must be given
by
\begin{eqnarray}
\delta B & = & 2\pi\beta\int_{0}^{\pi}d\theta\left[N^{\perp}\sin\left(\theta\right)\left(\frac{\partial_{r}H\delta F+\partial_{r}F\delta H}{\sqrt{F}}-\frac{\sqrt{F}\partial_{r}H\delta H}{H}\right.\right.\nonumber \\
 &  & \left.\left.+\frac{\sqrt{F}H\partial_{r}\phi\delta\phi}{2}+2\sqrt{F}\partial_{r}\delta H\right)-\partial_{r}\left(2N^{\perp}\sin\left(\theta\right)\sqrt{F}\right)\delta H-A_{\tau}\delta p^{r}\right]_{r_{+}}^{\infty}\nonumber \\
 &  & +2\pi\beta\int_{r+}^{\infty}dr\left[\frac{N^{\perp}e^{-\sqrt{3}\phi}}{H\sqrt{F}\sin\left(\theta\right)}\partial_{\theta}A_{\varphi}\delta A_{\varphi}\right]_{0}^{\pi}.\label{eq:deltaB}
\end{eqnarray}
In this expression is possible to recognize two kind of terms, the
surface term coming from a total derivative in the radial coordinate
and a  boundary term that comes from a total derivative in the polar
angle, which clearly is not vanishing since the presence of an angular
component depending on the polar angle in the gauge field. The analysis
of the variation of the term $B$ and the evaluation on the AdS$_{4}$
dyonic black hole \eqref{eq:deltaB} will be performed in the following
subsection.

\subsection{Gibbs free energy and first law}

From \eqref{eq:deltaB}, we can identify different contributions depending
whether the term comes from a total derivative in the radial coordinate,
or whether the term comes from a total derivative in the polar angle
which is identified as a polar boundary term. The surface term evaluated
at infinity will be denoted by $\delta B\left(\infty\right)$ and,
on the other hand, $\delta B(r_{+})$ will stand for the surface term
at the horizon. The polar boundary term in turn will be denoted by
$\delta B_{\theta}$. Hence, the variation of $B$ \eqref{eq:deltaB}
can be written as 
\begin{equation}
\delta B=\delta B(\infty)+\delta B(r_{+})+\delta B_{\theta},
\end{equation}
where the surface term at infinity $\delta B(\infty)$ is given by
\begin{eqnarray}
\delta B(\infty) & = & 2\pi\beta\int_{0}^{\pi}d\theta\left[N^{\perp}\sin\left(\theta\right)\left(\frac{\partial_{r}H\delta F+\partial_{r}F\delta H}{\sqrt{F}}-\frac{\sqrt{F}\partial_{r}H\delta H}{H}\right.\right.\nonumber \\
 &  & \left.\left.+\frac{\sqrt{F}H\partial_{r}\phi\delta\phi}{2}+2\sqrt{F}\partial_{r}\delta H\right)-\partial_{r}\left(2N^{\perp}\sin\left(\theta\right)\sqrt{F}\right)\delta H-A_{\tau}\delta p^{r}\right]_{\infty},\nonumber \\
\label{eq:dB-Infinity}
\end{eqnarray}
the surface term at the horizon $\delta B(r_{+})$ is
\begin{eqnarray}
\delta B(r_{+}) & = & -2\pi\beta\int_{0}^{\pi}d\theta\left[N^{\perp}\sin\left(\theta\right)\left(\frac{\partial_{r}H\delta F+\partial_{r}F\delta H}{\sqrt{F}}-\frac{\sqrt{F}\partial_{r}H\delta H}{H}\right.\right.\nonumber \\
 &  & \left.\left.+\frac{\sqrt{F}H\partial_{r}\phi\delta\phi}{2}+2\sqrt{F}\partial_{r}\delta H\right)-\partial_{r}\left(2N^{\perp}\sin\left(\theta\right)\sqrt{F}\right)\delta H-A_{\tau}\delta p^{r}\right]_{r_{+}},\nonumber \\
\label{eq:dB-Horizon}
\end{eqnarray}
and the polar boundary term $\delta B_{\theta}$ reads as
\begin{equation}
\delta B_{\theta}=2\pi\beta\int_{r+}^{\infty}dr\left[\frac{N^{\perp}e^{-\sqrt{3}\phi}}{H\sqrt{F}\sin\left(\theta\right)}\partial_{\theta}A_{\varphi}\delta A_{\varphi}\right]_{0}^{\pi}.\label{eq:dB-Bulk}
\end{equation}

Once the different contributions to the variation of $B$ are identified
one can analyze their physical content. In the surface term at infinity
$\delta B\left(\infty\right)$ is possible to find the variation of
the charges coming from symmetries of the action together with their
respective chemical potentials, which correspond to the Lagrange multipliers
at infinity of the respective symmetry (as was shown in Section \ref{Section3}).
This is because at the end of the day the term \eqref{eq:dB-Infinity}
is obtained from the boundary term of the Hamiltonian which ensures
that the canonical generators have well-defined functional derivatives
\cite{Regge:1974zd}. Then, from $\delta B\left(\infty\right)$ it
will be identified the variations of the mass and the electric charge
of the AdS$_{4}$ dyonic dilatonic black hole. From the surface term
at the horizon $\delta B\left(r_{+}\right)$ will be obtained the
entropy of the black hole which corresponds to the Bekenstein-Hawking
entropy. Finally, from the polar boundary term $\delta B_{\theta}$
can be identified the contribution of the topological charge of the
system leading to the variation of the magnetic charge multiplied
by the magnetic potential.

Let us to introduce the Euclidean continuation of the AdS$_{4}$ dyonic
dilatonic black hole that satisfies the field equations \eqref{eq:eq1}-\eqref{eq:eq5}
and the constraints \eqref{eq:Constraints}. This is obtained after
performing the identifications $t\rightarrow-i\tau$ and $\beta_{1}\rightarrow-\beta_{1}$
in the Lorentzian solution, which make the black hole functions to
take the form 
\begin{equation}
H_{1}\left(r\right)=\gamma_{1}^{-1}\left(1+2\beta_{1}f_{0}\left(r\right)-\beta_{1}\beta_{2}f_{0}\left(r\right)^{2}\right),\qquad H_{2}\left(r\right)=\gamma_{2}^{-1}\left(1-2\beta_{2}f_{0}\left(r\right)-\beta_{1}\beta_{2}f_{0}\left(r\right)^{2}\right),
\end{equation}
where $\gamma_{1}=1+2\beta_{1}-\beta_{1}\beta_{2}$ and $\gamma_{2}=1-2\beta_{2}-\beta_{1}\beta_{2}$.
The functions in the line element \eqref{eq:Line-Element-MS} are
considered to be
\begin{equation}
F\left(r\right)=\frac{f\left(r\right)}{\sqrt{H_{1}\left(r\right)H_{2}\left(r\right)}},\qquad H\left(r\right)=\sqrt{H_{1}\left(r\right)H_{2}\left(r\right)}r^{2},
\end{equation}
where the function $f\left(r\right)$ is the same as the one given
in \eqref{eq:function-f}. The lapse function is $N^{\perp}\left(r\right)=\sqrt{F\left(r\right)}$.
The scalar field is defined in \eqref{eq:Scalar-Field}, and the temporal
component of the gauge field is given by
\begin{equation}
A_{\tau}\left(r\right)=-\frac{\sqrt{2}\left(1-H_{1}\left(r\right)+\beta_{1}\left(f_{0}\left(r\right)-H_{1}\left(r\right)\right)\right)}{\sqrt{\beta_{1}\gamma_{2}}H_{1}\left(r\right)}+\Phi_{e}.
\end{equation}
Note that the possibility of adding a constant $\Phi_{e}$ allows
to have a regular gauge field at the horizon. This constant is related
to the electrostatic potential of the solution when the regularity
conditions on the black hole are established. The angular component
of the gauge field takes the same definition given in \eqref{eq:Aphi}.

Replacing the Euclidean continuation of the AdS$_{4}$ dyonic dilatonic
black hole in the surface term at infinity $\delta B\left(\infty\right)$
given in eq. \eqref{eq:dB-Infinity}, it is obtained that
\begin{equation}
\delta B\left(\infty\right)=-\beta\delta M-\beta\Phi_{e}\delta Q_{e},
\end{equation}
where the variations of the mass and electric charge are respectively
given by
\begin{eqnarray}
\delta M & = & \delta\left(\frac{16\pi\left(1+\beta_{1}\right)\left(1-\beta_{2}\right)\left(1+\beta_{1}\beta_{2}\right)\mu}{\gamma_{1}\gamma_{2}}\right)+\Theta,\label{eq:dM}\\
\delta Q_{e} & = & 4\pi\delta\left(\frac{2\sqrt{2}\mu\sqrt{\beta_{1}\gamma_{2}}}{\gamma_{1}}\right).\label{eq:dQe}
\end{eqnarray}
The above variations coincide with the values computed in \eqref{eq:deltaM}
and \eqref{eq:Q}. In the variation of the mass is clearly obtained
a contribution
\begin{eqnarray}
\Theta & = & \frac{64\pi g^{2}\mu^{3}\left(1+\beta_{1}\beta_{2}\right)\left(\beta_{1}+\beta_{2}\right)^{2}\gamma}{\gamma_{1}^{3}\gamma_{2}^{3}}+\Phi^{E}\nonumber \\
 & = & -\frac{32\pi g^{2}\mu^{3}\left(\beta_{1}+\beta_{2}\right)}{\gamma_{1}^{2}\gamma_{2}^{2}}\left(\beta_{2}\left(1-2\beta_{1}-2\beta_{2}+\beta_{1}\beta_{2}\right)\delta\beta_{1}-\beta_{1}\left(1+2\beta_{1}+2\beta_{2}+\beta_{1}\beta_{2}\right)\delta\beta_{2}\right),\nonumber \\
\end{eqnarray}
where $\Phi^{E}$ is the Euclidean continuation of $\Phi$. Here $\Theta$
corresponds to the supposed scalar charge term in the context of \cite{Lu:2013ura}. 

The inverse of the temperature $\beta$ and the electrostatic potential
$\Phi_{e}$ are determined through the regularity conditions at the
horizon. Indeed,
\begin{equation}
\beta=\frac{4\pi\sqrt{H_{1}\left(r_{+}\right)H_{2}\left(r_{+}\right)}}{f\mbox{\ensuremath{'}}\left(r_{+}\right)},\qquad\Phi_{e}=-\sqrt{\frac{2}{\beta_{1}\gamma_{2}}}\left(1+\beta_{1}-\frac{1+\beta_{1}f_{0}\left(r_{+}\right)}{H_{1}\left(r_{+}\right)}\right).\label{eq:ChemicalPotentials}
\end{equation}
The temperature is obtained by demanding the absence of conical singularities
around the event horizon, while the electrostatic potential comes
from the trivial holonomy condition of the gauge field around a temporal
cycle on the plane $r-\tau$ at the event horizon. Introducing the
values of the chemical potentials \eqref{eq:ChemicalPotentials} into
the surface term at the horizon $\delta B\left(r_{+}\right)$, we
get that this term exactly coincides with the Bekenstein-Hawking entropy
\begin{equation}
\delta B\left(r_{+}\right)=\delta\left(16\pi^{2}\sqrt{H_{1}\left(r_{+}\right)H_{2}\left(r_{+}\right)}r_{+}^{2}\right)=\delta S.\label{eq:dBrp}
\end{equation}
The polar boundary term $\delta B_{\theta}$ has to be carefully computed
by using the definition of the angular component of the gauge field
given in \eqref{eq:Aphi}. Then, 
\begin{eqnarray}
\delta B_{\theta} & = & 2\pi\beta\left(\int_{r+}^{\infty}dr\frac{e^{-\sqrt{3}\phi}}{H}\right)\left(\left[\frac{\partial_{\theta}A_{\varphi}\delta A_{\varphi}}{\sin\left(\theta\right)}\right]_{0}^{\pi/2-\delta}+\left[\frac{\partial_{\theta}A_{\varphi}\delta A_{\varphi}}{\sin\left(\theta\right)}\right]_{\pi/2+\delta}^{\pi}\right)_{\delta\rightarrow0}\nonumber \\
 & = & -2\pi\beta\left(\int_{r+}^{\infty}dr\frac{e^{-\sqrt{3}\phi}}{H}\right)\left(\left[p\delta p\left(1+\cos\left(\theta\right)\right)\right]_{0}^{\pi/2-\delta}+\left[p\delta p\left(-1+\cos\left(\theta\right)\right)\right]_{\pi/2+\delta}^{\pi}\right)_{\delta\rightarrow0}\nonumber \\
 & = & 4\pi\beta\left(\int_{r+}^{\infty}dr\frac{e^{-\sqrt{3}\phi}}{H}\right)p\delta p.\label{eq:dBtheta-1}
\end{eqnarray}
This term can be conveniently written as
\begin{equation}
\delta B_{\theta}=-\beta\Phi_{m}\delta Q_{m},\label{eq:dBtheta2}
\end{equation}
where it can be identified the magnetic potential
\begin{equation}
\Phi_{m}=-\sqrt{\frac{2}{\beta_{2}\gamma_{1}}}\left(1-\beta_{2}-\frac{1-\beta_{2}f_{0}\left(r_{+}\right)}{H_{2}\left(r_{+}\right)}\right),
\end{equation}
and also the value of variation of the magnetic charge
\begin{equation}
\delta Q_{m}=4\pi\delta\left(\frac{2\sqrt{2}\mu\sqrt{\beta_{2}\gamma_{1}}}{\gamma_{2}}\right).\label{eq:dQm}
\end{equation}
In consequence, the variation of the boundary term $B$ is given by
\begin{equation}
\delta B=\delta S-\beta\delta M-\beta\Phi_{e}\delta Q_{e}-\beta\Phi_{m}\delta Q_{m}.
\end{equation}
Note that once this term is integrated, the value of $B$ corresponds
to the Euclidean Hamiltonian action $I^{E}$ evaluated on stationary
configurations and on the constraints surface. In the grand canonical
ensemble $I^{E}$ is related to the Gibbs free energy as $I^{E}=-\beta G$.
It is also worth to point out that since the first law of black hole
thermodynamics,
\begin{equation}
\delta M=T\delta S-\Phi_{e}\delta Q_{e}-\Phi_{m}\delta Q_{m},\label{eq:First-Law}
\end{equation}
is a consequence that the Euclidean action to attain an extremum,
\eqref{eq:First-Law} is identically satisfied independently of the
boundary conditions on the mass. This is because \eqref{eq:First-Law}
is a relation that only involves the variation of the conserved charges.
This can be explicitly proved by introducing the computed value for
the charge variations \eqref{eq:dM}, \eqref{eq:dQe}, \eqref{eq:dQm}
and the chemical potentials obtained by using regularity conditions
\eqref{eq:ChemicalPotentials} into \eqref{eq:First-Law}.

Once the mass is integrated using arbitrary boundary conditions (see
Section \ref{Section3}), it is possible to find the value of the
Gibbs free energy which is equivalent Euclidean Hamiltonian action
evaluated on-shell,
\begin{equation}
I^{E}=S-\beta M-\beta\Phi_{e}Q_{e}-\beta\Phi_{m}Q_{m}.
\end{equation}
Recalling that we have chosen the grand canonical ensemble and taking
the Euclidean action as our thermodynamic potential, the values
of the extensive quantities; mass, electric charge, magnetic charge
and the entropy are obtained through the following thermodynamic
relations
\begin{eqnarray}
M & = & -\left(\frac{\partial I^{E}}{\partial\beta}\right)_{\Phi_{e},\Phi_{m}}+\frac{\Phi_{e}}{\beta}\left(\frac{\partial I^{E}}{\partial\Phi_{e}}\right)_{\beta,\Phi_{m}}+\frac{\Phi_{m}}{\beta}\left(\frac{\partial I^{E}}{\partial\Phi_{m}}\right)_{\beta,\Phi_{e}},\\
Q_{e} & = & -\frac{1}{\beta}\left(\frac{\partial I^{E}}{\partial\Phi_{e}}\right)_{\beta,\Phi_{m}},\\
Q_{m} & = & -\frac{1}{\beta}\left(\frac{\partial I^{E}}{\partial\Phi_{m}}\right)_{\beta,\Phi_{e}},\\
S & = & I_{E}-\beta\left(\frac{\partial I^{E}}{\partial\beta}\right)_{\Phi_{e},\Phi_{m}}.
\end{eqnarray}
The values of the charges and the entropy computed above coincide
with \eqref{eq:Masss}, \eqref{eq:Q}, \eqref{eq:dQm} and \eqref{eq:dBrp},
respectively.

\section{Concluding remarks \label{Section5}}

We have carried out the thermodynamic analysis of a new class of AdS$_{4}$
dyonic dilatonic black hole recently proposed in \cite{Lu:2013ura},
which is a solution of the bosonic sector of a Kaluza-Klein dimensional
reduction of eleven-dimensional supergravity. The Noether conserved
charges were computed using the Regge-Teitelboim Hamiltonian approach.
These correspond to the mass, which acquires contributions from the
scalar field and the electric charge. It was also shown that the
mass acquires non-integrable contributions from the scalar field,
then it was necessary to impose integrability conditions for having
a definite mass. These conditions are generically solved by introducing
boundary conditions that arbitrarily relate the leading and subleading
terms of the scalar field fall-off. A possible physical criterium
to establish the arbitrary functions coming from the integrability
condition is to preserve the AdS symmetry of the scalar field behavior
at infinity as was established in \cite{Hertog:2004dr}, \cite{Henneaux:2006hk}.
The Hamiltonian Euclidean action was computed by demanding the action
to attain an extremum, where its value was given by the corresponding
radial boundary term plus an additional polar
angle boundary term, since the presence of a magnetic monopole. The
computation was performed in the grand canonical ensemble. From the
thermodynamic analysis, the conserved charges were identified. Noether
charges, mass and electric charge, were obtained from the radial boundary
term at infinity unlike the magnetic charge, that comes from the additional
polar angle boundary term. Remarkably, the magnetic potential appeared
already determined in the variation of such boundary term, unlike
the chemical potentials associated to Noether charges which are Lagrange
multipliers of the system at infinity. They are obtained through regularity
conditions at the horizon. Considering the above, it is possible to
verify that the first law of black hole thermodynamics is identically
satisfied. This is a consequence of having a well-defined and finite
Hamiltonian action principle. 

A different way to deal with the thermodynamics of dyonic black holes
is to consider a manifestly duality invariant action that involves
two U(1) symmetries, producing the appearance of electric and magnetic
Gauss constraints \cite{Barnich:2007uu}. The dyonic Reissner-Nordström
black hole is solution of system proposed in \cite{Barnich:2007uu},
however in this case the magnetic and electric field appear as Coulombian
potentials, hence the solution is devoid of stringy singularities.
In this case, all the conserved charges appearing in the first law
come from symmetries of the action. 

It would be interesting to analyze the existence of phase transitions
between this dyonic dilatonic black hole solution and the dyonic Reissner-Nordström
black hole, and studying the chance that bellow a critical temperature
the dyonic Reissner-Nordström black hole undergoes spontaneously to
a dressing state with a dilaton scalar field. This kind of results
have been reproduced, for instance, in the case of four-dimensional
topological black holes dressed with a scalar field in \cite{Martinez:2010ti}. 

\acknowledgments We thank Hernán A. González, Cristián Martínez,
Alfredo Pérez, David Tempo, Ricardo Troncoso and Jorge Zanelli for
useful comments, and specially to Nathalie Deruelle for encouraging
us to address this problem and also for enlightening and fruitful
discussions. M.C. thanks Conicyt for financial support. This research
has been partially supported by Fondecyt grant 3150448. The Centro
de Estudios Científicos (CECs) is funded by the Chilean Government
through the Centers of Excellence Base Financing Program of Conicyt.


\begin{thebibliography}{10}
\bibitem{Lu:2013ura}H.~L\"{u}, Y.~Pang and C.~N.~Pope,   ``AdS Dyonic Black Hole and its Thermodynamics,''   JHEP {\bf 1311}, 033 (2013)  [arXiv:1307.6243 [hep-th]].   

\bibitem{Lu:2014maa}H.~L\"{u}, C.~N.~Pope and Q.~Wen,   ``Thermodynamics of AdS Black Holes in Einstein-Scalar Gravity,''   JHEP {\bf 1503}, 165 (2015)      [arXiv:1408.1514 [hep-th]].   

\bibitem{Cvetic:1999xp}M.~Cveti\v{c} {\it et al.},   ``Embedding AdS black holes in ten-dimensions and eleven-dimensions,''   Nucl.\ Phys.\ B {\bf 558}, 96 (1999)      [hep-th/9903214].   

\bibitem{Henneaux:2006hk}M.~Henneaux, C.~Mart\'inez, R.~Troncoso and J.~Zanelli,   ``Asymptotic behavior and Hamiltonian analysis of anti-de Sitter gravity coupled to scalar fields,''   Annals Phys.\  {\bf 322}, 824 (2007) [hep-th/0603185]

\bibitem{Barnich:2002pi}G.~Barnich,   ``Conserved charges in gravitational theories: Contribution from scalar fields,''   Ann.\ U.\ Craiova Phys.\  {\bf 12}, no. III, 14 (2002)   [gr-qc/0211031].   

\bibitem{Compere:2009dp}G.~Comp\`{e}re, K.~Murata and T.~Nishioka,   ``Central Charges in Extreme Black Hole/CFT Correspondence,''   JHEP {\bf 0905}, 077 (2009)      [arXiv:0902.1001 [hep-th]].   

\bibitem{Cardenas:2014kaa}M.~C\'ardenas, O.~Fuentealba and C.~Mart\'inez,   ``Three-dimensional black holes with conformally coupled scalar and gauge fields,''   Phys.\ Rev.\ D {\bf 90}, no. 12, 124072 (2014)      [arXiv:1408.1401 [hep-th]].   

\bibitem{Chow:2013gba}D.~D.~K.~Chow and G.~Comp\`{e}re,   ``Dyonic AdS black holes in maximal gauged supergravity,''   Phys.\ Rev.\ D {\bf 89}, no. 6, 065003 (2014)      [arXiv:1311.1204 [hep-th]].   

\bibitem{Hertog:2004dr}T.~Hertog and K.~Maeda,   ``Black holes with scalar hair and asymptotics in N = 8 supergravity,''   JHEP {\bf 0407}, 051 (2004)      [hep-th/0404261].   

\bibitem{deHaro:2000vlm}S.~de Haro, S.~N.~Solodukhin and K.~Skenderis,   ``Holographic reconstruction of space-time and renormalization in the AdS / CFT correspondence,''   Commun.\ Math.\ Phys.\  {\bf 217}, 595 (2001)     [hep-th/0002230].   

\bibitem{Dirac:1948um}P.~A.~M.~Dirac,   ``The Theory of magnetic poles,''   Phys.\ Rev.\  {\bf 74}, 817 (1948).     

\bibitem{Wu:1976qk}T.~T.~Wu and C.~N.~Yang,   ``Dirac's Monopole Without Strings: Classical Lagrangian Theory,''   Phys.\ Rev.\ D {\bf 14}, 437 (1976).      

\bibitem{Wu:1976ge}T.~T.~Wu and C.~N.~Yang,   ``Dirac Monopole Without Strings: Monopole Harmonics,''   Nucl.\ Phys.\ B {\bf 107}, 365 (1976).    

\bibitem{Regge:1974zd}T.~Regge and C.~Teitelboim,   ``Role of Surface Integrals in the Hamiltonian Formulation of General Relativity,''   Annals Phys.\  {\bf 88}, 286 (1974).      

\bibitem{Ashtekar:1984zz}A.~Ashtekar and A.~Magnon,   ``Asymptotically anti-de Sitter space-times,''   Class.\ Quant.\ Grav.\  {\bf 1}, L39 (1984).    

\bibitem{Ashtekar:1999jx}A.~Ashtekar and S.~Das,   ``Asymptotically Anti-de Sitter space-times: Conserved quantities,''   Class.\ Quant.\ Grav.\  {\bf 17}, L17 (2000)     [hep-th/9911230].   

\bibitem{Anabalon:2014fla}A.~Anabal\`{o}n, D.~Astefanesei and C.~Mart\`{i}nez, ``Mass of asymptotically anti-de Sitter hairy spacetimes,'' Phys.\ Rev.\ D {\bf 91}, no. 4, 041501 (2015) [arXiv:1407.3296[hep-th]]. 

\bibitem{Barnich:2007uu}G.~Barnich and A.~Gomberoff,   ``Dyons with potentials: Duality and black hole thermodynamics,''   Phys.\ Rev.\ D {\bf 78}, 025025 (2008)      [arXiv:0705.0632 [hep-th]].   

\bibitem{Martinez:2010ti}C.~Mart\'{i}nez and A.~Montecinos,   ``Phase transitions in charged topological black holes dressed with a scalar hair,''   Phys.\ Rev.\ D {\bf 82}, 127501 (2010)      [arXiv:1009.5681 [hep-th]].   
\end{thebibliography}
\end{document}